\documentstyle[preprint,tighten,aps,epsfig]{revtex}

\newcommand{\tr}{\mbox{Tr}\,}
\newcommand{\egmm}{\eta\rightarrow\gamma\mu^+\mu^-}
\begin{document}

\preprint{
\begin{tabular}{r} FTUV/98$-$25 \\ IFIC/98$-$25
\end{tabular}
}

\title{$\eta\gamma Z$ anomaly from the $\egmm$ decay}
\author{J.\ Bernab\'eu, D.\ G\'omez Dumm 
and J.\ Vidal}
\address{\hfill \\ Departament de F\'{\i}sica Te\`orica, IFIC, 
CSIC -- Universitat de Val\`encia \\
Dr.\ Moliner 50, E-46100 Burjassot (Val\`encia), Spain}

\maketitle

\begin{abstract}
We show that the $\eta\gamma Z$ anomaly can be measured by
analysing parity-violating effects in the $\egmm$ decay. In this sense,
we find that the longitudinal polarization of the outgoing $\mu^+$ is
an appropriate observable to be considered in future high-statistics
$\eta$ factories. The effect is expected to lie in the range
$10^{-5}-10^{-6}$ in the Standard Model.
\end{abstract}



\vspace{2cm}

The production of $\eta$ mesons with high statistics in future experiments
will improve the present knowledge of rare $\eta$ decay processes, enabling
new possible tests of the Standard Model predictions. In particular, a
sufficiently large number of events would allow the experimental
observation of weak interaction effects, which are in general hardly
suppressed by the large mass of the $W^\pm$ and $Z$ gauge bosons.
Parity-violating observables are obvious candidates to get relevant
information in this sense.

Among the different $\eta$ decay channels, let us focus our attention
on those which involve the effective coupling $\eta\gamma\gamma$, with
$\gamma$ either a real or virtual photon. These processes deserve a
significant theoretical interest, since the $\eta\gamma\gamma$ vertex
is governed by the axial anomaly, i.e., it provides a direct evidence of
the presence of quantum corrections breaking the $U(1)_A$ symmetry of the
QCD Lagrangian \cite{adl}. A consistent description for these decays
can be obtained within the framework of Chiral Perturbation Theory (ChPT),
where the anomaly is introduced through the Wess-Zumino-Witten (WZW)
functional \cite{wzw}. As it is well known, this leads also to
successful predictions for processes involving a $\pi\gamma\gamma$
coupling, such as the decays $\pi\rightarrow\gamma\gamma$
(where the anomaly was actually discovered) or $\pi\rightarrow\gamma
e^+ e^-$ \cite{bij}.

Now, if weak interactions are taken into account, the presence of an
anomalous effective vertex $\eta\gamma Z$ is expected as well. The latter
should be correctly described in an analogous way as the $\eta\gamma\gamma$
one, the quark electromagnetic couplings being replaced
by the vector part of the corresponding weak neutral currents. We point
out that this ``$Z$ anomaly'' has never been
measured experimentally. Clearly, in order
to get an observable effect, it would be necessary to search for an
asymmetry that could disentangle the $Z$ contribution. In this letter, we
show that the $\egmm$ channel is an appropriate one for this purpose. Indeed,
owing to the parity-violating nature of the weak interactions, this process
offers the possibility of constructing
the required asymmetry by looking at the polarization
of one of the final muons. We perform here an explicit calculation of the
longitudinal and transverse polarizations of the $\mu^+$ (the easiest to be
measured, in view of the $\mu^-$ capture produced in the polarimeters), and
give a numerical estimate of the effects that can be expected.

Our analysis will be carried out within the framework of ChPT. First of all,
let us remark that the treatment of the $\eta\gamma\gamma$ anomaly is not
completely equivalent to the $\pi\gamma\gamma$ one: there is an additional
difficulty arising from the so-called $\eta-\eta'$ mixing. As it is well
known, in the isospin limit the $\eta$ mass state is given in general by a
mixing between the $I=0$ states $\eta_8$ and $\eta_0$, octet and singlet
respectively under chiral $SU(3)$. The problem is that, due to the
presence of the axial anomaly,
the singlet state $\eta_0$ cannot (in principle) be treated as an
approximate Goldstone boson of the theory, and consequently its interactions
are not described by ChPT. Still, however, it is possible to take into
account the approximation of large number of colours. It can be shown that
in the large $N_c$ limit,
the $U(1)_A$ symmetry of the Lagrangian is restored at the quantum level,
and the $\eta_0$ field is indeed incorporated as a ninth Goldstone boson
\cite{wit}. In this way, it is possible to get definite predictions for the
interactions of the $\eta_0$ by performing a double expansion in momenta
and $N_c^{-1}$. This is the procedure we will follow in this work.
In fact, when looking at the $\eta\gamma Z$ anomaly, it is found that the
contribution of the $\eta_8$ is significantly suppressed, so that the
$\eta_0$ part turns out to be the dominant one. This means that the
measurement of the observables proposed here would represent an important
test not only for the $Z$ anomaly itself, but also for the viability of the
large $N_c$ approximation.

\hfill

Let us concentrate on the $\eta\rightarrow\gamma\mu^+\mu^-$
decay channel. We begin by writing down the squared amplitude
for the process, which is represented by the diagram in Fig.\ 1. One has
in general
\begin{equation}
{\cal M}=i\epsilon^{\rho\nu\alpha\beta}q_\alpha k_\beta 
\varepsilon_\rho^{(\lambda)\ast} \left[ \frac{C_\gamma}{q^2}\, f(q^2)\, 
e\;\bar u(p^-) \gamma_\nu v(p^+) + \frac{C_{\gamma Z}}{M_Z^2}\, 
\frac{g}{2 \cos \theta_W} \;\bar u(p^-) \gamma_\nu 
(g_V^l + g_A^l \gamma_5) v(p^+) \right]\, ,
\label{msq} 
\end{equation} 
where $\epsilon_\rho^{(\lambda)}$ is the photon polarization four-vector,
$g_V^l$ and $g_A^l$ are the lepton weak neutral couplings, and
$C_\gamma$ and $C_{\gamma Z}$ stand for the
anomalous vertices $\eta\gamma\gamma^\ast$ and $\eta\gamma Z^\ast$,
respectively.
Notice that we have included a form factor $f(q^2)$ for the off-shell photon;
we will make use of a single-pole approximation, taking
\begin{equation}
f(q^2)=\left( 1-\frac{q^2}{\Lambda^2}\right)^{-1} \, .
\label{ffac}
\end{equation}
An experimental fit of the slope parameter $\Lambda^{-2}$ has been done some
time ago \cite{fq1} resulting in $\Lambda=720\pm 90$ MeV, in good
agreement with the hypothesis of $\rho$ meson dominance\footnote{In fact, the
averaged slope $\Lambda^{-2}$ turns out to be  
slightly smaller ($\sim 1/m_\rho^2$) when data from $\gamma \gamma^\ast 
\rightarrow \eta$ processes are also included \cite{fq2}. However, these 
measurements have been performed at relatively large $(-q^2)$ values and
require an extrapolation. We keep here the result of Ref.\ \cite{fq1}, which
was taken directly from the $\eta\rightarrow\gamma \mu^+\mu^-$ decay.}.
 
In order to evaluate the anomalous vertices, let us first separate the
contributions of the $\eta_0$ and $\eta_8$ states. As mentioned above, the
mass eigenstate $\eta$ is given in general by a mixing
\begin{equation}
|\eta\rangle = \cos\theta_P |\eta_8\rangle - \sin\theta_P |\eta_0\rangle \,,
\label{mix}
\end{equation}
where the angle $\theta_P$ is a parameter that can be estimated either
through the
diagonalization of the $\eta$-$\eta'$ mass matrix, or by analysing the
phenomenology of $\eta$ and $\eta'$ decays. Both procedures are consistent
in ChPT at one-loop order, leading to a value of $\theta_P$ of about
$-20^\circ$ \cite{gk}. In this way, the couplings $C_\gamma$ and
$C_{\gamma Z}$ in (\ref{msq}) can be conveniently written as
\begin{equation}
C_{\gamma,\gamma Z} = C^{(8)}_{\gamma,\gamma Z} \cos\theta_P - 
C^{(0)}_{\gamma,\gamma Z} \sin\theta_P \,.
\label{mezcla}
\end{equation} 
The values of $C^{(8)}_\gamma$ and $C^{(8)}_{\gamma Z}$, i.e., those which
correspond to the octet state, can be easily obtained from the WZW
effective Lagrangian. One has
\begin{eqnarray} 
C^{(8)}_\gamma & = & \frac{N_c \,\alpha}{\pi f_{\eta_8}} 
\,\tr \left[ Q^2 \lambda_8 \right] = \frac{1}{\sqrt{3}}\,  
\frac{\alpha}{\pi f_{\eta_8}} \nonumber \\ 
C^{(8)}_{\gamma Z} & = & \frac{N_c \, e g}{8\pi^2 f_{\eta_8} 
\cos\theta_W} \,\tr \left[ Q g_V \lambda_8 \right] = \frac{1}{\sqrt{3}}\, 
\frac{e\,g}{16\pi^2 f_{\eta_8}\cos\theta_W} \,(1-4 \sin^2\theta_W) \,,
\label{c8} 
\end{eqnarray} 
where $Q$ and $g_V$ are defined as $\mbox{diag}(Q^u,Q^d,Q^s)$ and
$\mbox{diag}(g_V^u,g_V^d,g_V^s)$ respectively, and the parameter
$f_{\eta_8}$ is equal to the pion decay constant $f_\pi$ in the 
chiral limit. Notice that $C^{(8)}_{\gamma Z}$ is found
to be suppressed by a factor $(1-4\sin^2\theta_W)$ \cite{gab},
as it is demanded by the lack of anomalies in the SM: quark and
lepton contributions have to amount to the same magnitude and opposite
sign.

The evaluation of $C_{\gamma, \gamma Z}^{(0)}$ is more subtle. As stated
above, since $\eta_0$ is not a Goldstone boson in the chiral limit,
its couplings are in principle not described by ChPT. However, we can
take into account the large $N_c$ limit in order to get analogous
expressions to those in (\ref{c8}). At leading order in $N_c^{-1}$,
the chiral symmetry is enlarged to $U(3)$, and the WZW Lagrangian can be
extended to incorporate the $\eta_0$ field. One gets in this way
\begin{eqnarray} 
C^{(0)}_\gamma & = & \frac{\sqrt{2}}{\sqrt{3}}\,
\frac{N_c \,\alpha}{\pi f_{\eta_0}} 
\,\tr \left[ Q^2 \right] = \frac{2\sqrt{2}}{\sqrt{3}}\,  
\frac{\alpha}{\pi f_{\eta_0}} \nonumber\\ 
C^{(0)}_{\gamma Z} & = & \frac{\sqrt{2}}{\sqrt{3}}\,
\frac{N_c \, e g}{8\pi^2 f_{\eta_0}  
\cos\theta_W} \,\tr \left[ Q g_V \right] = \frac{\sqrt{2}}{\sqrt{3}} 
\,\frac{e\,g}{4\pi^2 f_{\eta_0}\cos\theta_W}\, (1-2 \sin^2\theta_W) \,,
\label{c0} 
\end{eqnarray}
where once again the relation $f_{\eta_0}=f_\pi$ is expected to hold at
the lowest order in the chiral expansion. In fact, if $f_{\eta_8}$ is
identified with the axial current decay constant corresponding to the
$\eta_8$ state, one finds at next to leading order (NLO) in ChPT \cite{f8rel}
\begin{equation}
f_{\eta_8}\simeq 1.3 f_\pi\,,
\label{efe8}
\end{equation} 
and then, from the experimental value of the $\eta'\rightarrow\gamma\gamma$ 
decay,
\begin{equation} 
f_{\eta_0}\simeq 1.1 f_\pi \,.
\label{efe0} 
\end{equation} 
It can be seen \cite{f0rel} that this value shows a very good agreement with
the NLO prediction given by ChPT in the $U(3)$ symmetric limit, thus
giving important support to the large $N_c$ approximation.

It is worth to notice from (\ref{c0}) that the $(1-4\sin^2\theta_W)$
suppression factor is not present in the case of $C^{(0)}_{\gamma Z}$.
In fact, for a mixing angle $\theta_P\simeq -20^\circ$, we see that the
$\eta_0$ state contribution to $C_{\gamma Z}$ is enhanced by about
a factor 10 with respect to that of the $\eta_8$, and largely dominates the
$\eta\gamma Z$ anomalous coupling.

\hfill

We proceed now to identify an observable that could be sensitive to the
$Z$ anomaly. As stated, in order to disentangle the $Z^\ast$
contribution to the $\egmm$ amplitude,
one is led to search for a parity-violating asymmetry.

We will consider two possible candidates, namely the
longitudinal and transverse polarizations of the final $\mu^+$ resulting
from the decay. Let us recall the amplitude in (\ref{msq}), and
perform the sum over spins and helicities for the $\mu^-$ and photon final
states respectively. We find
\begin{eqnarray} 
\sum_{s(\mu^-),\lambda} |{\cal M}|^2 & = & \frac{\left| B_V\right|^2}{q^4}
\left( f(q^2)\right)^2\left[ 2 q^2 \left( (q\cdot k)^2-2 (p^+\cdot k)
\, (p^-\cdot k) \right) + 4\, m_\mu^2 (q\cdot k)^2 \right] -
\frac{\mbox{Re} (B_V B_A^\ast)}{q^2} \nonumber \\
& & \hspace{-1cm} \times f(q^2)\left\{8\, m_\mu\left[(q\cdot k)\, 
(s\cdot p^-)\, (p^-\cdot k) + \frac{q^2}{2} (q\cdot k)\, (s\cdot k) 
- q^2 (s\cdot k)\, (p^-\cdot k)\right]\right\} \,,
\label{msum}
\end{eqnarray} 
where $s^\alpha=(s^0,\vec s)$ stands for the $\mu^+$ polarization
four-vector, and $B_V$ and $B_A$ correspond respectively to the vector and
axial vector muon couplings in (\ref{msq}) (we have assumed
$|B_A|^2 \ll |B_V|^2$). Considering the mixing in Eq.\
(\ref{mezcla}), we have
\begin{eqnarray}
B_V & = & e\, C_{\gamma}^{(8)}\left(\cos\theta_P - \frac{C_{\gamma}^{(0)}}{ 
C_{\gamma}^{(8)}} \,\sin\theta_P) \right) - \frac{g\, g_V^l}{2 \cos\theta_W} 
\,\frac{q^2}{M_Z^2} C_{\gamma Z}^{(8)} 
\left(\cos\theta_P - \frac{C_{\gamma Z}^{(0)}}{ 
C_{\gamma Z}^{(8)}}\,\sin\theta_P\right) \nonumber\\ 
B_A & = & - \frac{g\, g_A^l}{2 \cos\theta_W}\,
\frac{C_{\gamma Z}^{(8)}}{M_Z^2} 
\left( \cos\theta_P - \frac{C_{\gamma Z}^{(0)}}{ 
C_{\gamma Z}^{(8)}}\,\sin\theta_P \right)\; ,
\end{eqnarray}
and then, from relations (\ref{c8}) and (\ref{c0}),
\begin{eqnarray}
|B_V|^2 & \simeq & \frac{4 \alpha^3}{3\pi f_{\eta_8}^2} \,
\left( \cos\theta_P -2\sqrt{2} \frac{f_{\eta_8}}{f_{\eta_0}} \sin\theta_P
\right)^2 \nonumber\\
\mbox{Re} (B_V B_A^\ast) & \simeq & \frac{G_F}{\sqrt{2}}\,
\frac{\alpha^2}{6\pi^2 f_{\eta_8}^2} \,\left( \cos\theta_P
-2\sqrt{2} \frac{f_{\eta_8}}{f_{\eta_0}} \sin\theta_P \right) \nonumber\\
& &  \times \left( (1-4\sin^2\theta_W)\,\cos\theta_P - 4\sqrt{2}
\frac{f_{\eta_8}}{f_{\eta_0}} (1-2\sin^2\theta_W) \,\sin\theta_P \right)\,,
\label{ab}
\end{eqnarray} 
where we have kept only leading terms in powers of the weak effective
coupling $G_F$.

In order to deal with the phase space, we will define our
observables in the $\eta$ rest frame\footnote{Notice that the longitudinal
and transverse polarizations of the outgoing particles are in general not
invariant under Lorentz transformations. In fact, some small
dilution of the effect can be expected when the $\eta$ mesons are produced
in flight.}. Let us choose the $z$ axis along the $\mu^+$ three-momentum,
and take as independent variables the $\mu^+$ energy E and the angles
$\theta$ and $\varphi$ determining the direction of the outgoing photon.
For each differential phase space volume $d\Phi\equiv dE\,d\Omega$
(with $d\Omega=d\cos\theta\,d\varphi$), it is possible to define the
polarization of the $\mu^+$ along a given direction $\hat s$ by
\begin{equation}
P(E,\theta,\varphi;s)\equiv
\frac{d\Gamma^{(+)}/d\Phi - d\Gamma^{(-)}/d\Phi}{d\Gamma^{(+)}/d\Phi
+ d\Gamma^{(-)}/d\Phi}\,,
\label{difpol}
\end{equation}
where $\Gamma^{(\pm)}\equiv \Gamma(\pm s)$ are the widths to final states
with opposite $\mu^+$ polarization vectors. This observable is clearly
parity-violating, hence it will be dominated by the $\gamma^\ast-Z^\ast$
interference term in the squared amplitude. From Eq.\ (\ref{msum}),
the numerator in (\ref{difpol}) explicitly reads
\begin{equation}
\frac{d\Gamma^{(+)}-d\Gamma^{(-)}}{dE\,d\Omega} =
\frac{\mbox{Re}(B_V B_A^\ast)}{128 \pi^4} \frac{(m_\eta-2E)\; |\vec P|
\; f(q^2)}{m_\eta (2k^0-m_\eta)\,(m_\eta-E+|\vec P|\cos\theta)^2}
\times {\cal F}(E,\cos\theta;s)\,,
\label{dif}
\end{equation}
where the function ${\cal F}$ corresponds to the expression
in curly brackets in (\ref{msum}), and $k^0$ and
$\vec P$ stand for the photon energy and the $\mu^+$ three-momentum
respectively; in terms of $E$ and $\theta$,
\begin{equation}
q^2=m_\eta\,(m_\eta-2k^0)\;,\quad\quad
k^0=\frac{m_\eta\,(m_\eta/2-E)}{m_\eta-E+|\vec P|\cos\theta}\;,\quad 
\quad |\vec P|=\sqrt{E^2-m_\mu^2} \;.
\end{equation}
The form of ${\cal F}$ depends on the chosen direction of $\vec s$. In
the longitudinal case ($\vec s=E\vec P/m_\mu|\vec P|$), one has
\begin{equation}
\!{\cal F}_L = \frac{8\, m_\eta^3}{|\vec P|}
\left[ E \left( \frac{m_\eta}{2}-k^0\right)\!
\left({k^0}^2\!+(m_\eta-2E)\,(\frac{m_\eta}{2}
-E-k^0)\right)- \frac{m_\mu^2}{m_\eta}\, {k^0}^2(m_\eta-E-k^0)\right]
\label{fl}
\end{equation}
whereas for a transverse $\vec s$, we find
\begin{equation}
{\cal F}_T=8\,m_\mu\, m_\eta^2 k^0
\left[ \frac{m_\eta^2}{2} - (m_\eta-k^0)\,(E+k^0) \right]\,\sin\theta
\label{ft}
\end{equation}
(here, it is understood that $\vec s$ is oriented within the decay
plane, hence its direction is determined by the angle $\varphi$). On the
other hand, notice that the normal polarization (this means, normal to
the decay plane) is expected to be very small in this scenario, since it
is related to CP- or T-odd effects.

The denominator in (\ref{difpol}) is nothing but the differential
width for the $\egmm$ process. In this case the contribution of the
virtual $Z$ can be safely neglected, and Eq.\ (\ref{msum}) leads to
\begin{equation}
\frac{d\Gamma^{(+)}+d\Gamma^{(-)}}{dE\,d\Omega} \simeq
\,\frac{|B_V|^2}{128 \pi^4}\, \frac{(m_\eta-2E)\;
|\vec P|\;\left(f(q^2)\right)^2}{ [ m_\eta\,
(m_\eta-2k^0)\,(m_\eta-E+|\vec P|\cos\theta)]^2}
\times {\cal F}_0 (E,\cos\theta) \,,
\label{gam}
\end{equation}
with
\begin{equation}
{\cal F}_0 = 4\, m_\eta^2  
\left[ m_\eta \left( \frac{m_\eta}{2}-k^0\right) 
\left({k^0}^2+(m_\eta-2E)\,(\frac{m_\eta}{2} 
-E-k^0)\right)+m_\mu^2{k^0}^2\right] \,.
\label{fgam} 
\end{equation} 
By integrating the expression in (\ref{gam}) over the whole phase space,
one obtains a prediction for the
total width that can be compared with the experimental results. Using the 
value of $|B_V|^2$ in (\ref{ab}), a mixing angle $\theta_P\simeq -20^\circ$,
and taking $f_{\eta_8}$ and $f_{\eta_0}$ as in (\ref{efe8})
and (\ref{efe0}) respectively, we find
\begin{equation}
\Gamma(\eta\rightarrow\gamma\mu^+\mu^-)\simeq 3.6\times 10^{-7}
\mbox{ MeV} \,,
\end{equation} 
in good agreement with the value of $(3.7\pm 0.6)\times 10^{-7}$ MeV from 
the Particle Data Group~\cite{tabla}. 

Now, from (\ref{dif}) and (\ref{gam}), the asymmetry defined in Eq.\
(\ref{difpol}) is given by
\begin{equation}
P(E,\theta,\varphi;s)=-\frac{q^2\,\mbox{Re}(B_V B_A^\ast)}{f(q^2)\,
|B_V|^2}\,\frac{{\cal F}(E,\cos\theta;s)}{{\cal F}_0(E,\cos\theta)}\,.
\end{equation}
As expected, one finds here a strong suppression factor, arising from the
ratio between the $Z^\ast$ and $\gamma^\ast$ contributions to the decay
amplitude. A rough estimate of the order of magnitude for the effect
yields $|P|\sim m_\eta^2 |B_A/B_V| \sim 10^{-5}-10^{-6}$.

Let us finally perform a more detailed numerical analysis, considering
the expected $\mu^+$ polarization for a finite region
$\Delta\Phi$ of the phase space. In analogy with (\ref{difpol}), it is
possible to define the asymmetry
\begin{equation}
P(\Delta\Phi;s)\equiv\frac{\int d\Gamma^{(+)}- \int d\Gamma^{(-)}}{
\int d\Gamma^{(+)}+\int d\Gamma^{(-)}}\;,
\label{asim}
\end{equation}
where the integrals extend to the volume $\Delta\Phi$. We will concentrate
on the longitudinal $\mu^+$ polarization, looking at the dependence of both
numerator and denominator in (\ref{asim}) with the variables $E$ and
$\theta$ introduced above (the integration in $\varphi$ is trivial). In
general, we expect the value of $P$ to be optimized by choosing a convenient
region of the phase space. By
analysing the expression in (\ref{gam}), it can be seen that
the differential decay width is sharply peaked backwards (i.e., when the
photons are produced with opposite direction to that of the $\mu^+$),
with more than 70\% of the events in the $-1\leq\cos\theta\leq -0.5$
region. Unfortunately, the numerator in Eq.\ (\ref{asim}) shows a similar
behaviour, and we cannot get a significant enhancement in $|P|$ by
introducing a cut in $\cos\theta$. On the other hand, the $\mu^+$ energy
spectrum is shown in Fig.\ 2. The plotted curves result from the functions
in Eqs.\ (\ref{gam}) (solid) and (\ref{dif}) (dashed), after
integrating over all possible directions of the final photon; as
anticipated, the difference
between the rates to opposite $\mu^+$ polarizations is about six orders
of magnitude lower than the total decay width (notice the different scales
at both sides in Fig.\ 2). By looking at the figure, it is seen that the
value of $|P|$ can be
increased by performing a lower cut in the $\mu^+$ energy range. Indeed, a
convenient region is that given by 140 MeV $\alt E\leq m_\eta/2$,
in which we find $P\simeq -2.4\times 10^{-6}$, with 80\% of the
total number of events. Though it is still possible to obtain higher values
of $P$ by moving the cut towards the upper limit of $E$, the growth is
found to be slow in comparison with the reduction of statistics
(e.g. for $E\geq 220$ MeV, we get $P\simeq -3.8\times 10^{-6}$, while
only 20\% of the events remain).

The transverse $\mu^+$ polarization is less favoured from the experimental
point of view. It can be seen from Eq.\ (\ref{ft}) that the asymmetry
presents in this case an additional $m_\mu/m_\eta$ suppression, which
is indeed expected from chirality arguments (no transverse polarization is
obtained in the limit of vanishing muon mass). Moreover, for each event,
the observable requires the identification of the decay plane, which
defines the direction of the polarization vector. The values of $|P|$
obtained in this case fall typically in the range $10^{-6}-10^{-7}$.

\hfill

Summarizing, we have analysed here the $Z$ contribution to the decay
$\egmm$. We have shown that this channel can be an appropriate one
to find an observable effect of the anomalous coupling
$\eta\gamma Z$, which has never been measured experimentally up to
now. Our analysis has been performed using ChPT, together with large-$N_c$
considerations. This framework allows to deal with the
interactions involving not only the $\eta_8$ but also the $\eta_0$
component of the $\eta$ mass eigenstate. In fact, it
turns out that the
$\eta_0$ part is that which dominates the $\eta\gamma Z$ anomalous vertex.
In order to disentangle the contribution of the $Z$ boson to the decay
amplitude, we have considered the polarization of the final muons, which
give rise to parity-violating effects. In particular, the longitudinal
polarization of the $\mu^+$ is shown to be an adequate candidate for
the measurement of the $Z$ anomaly in future $\eta$-factory experiments.
The value of this observable in the $\eta$ rest frame is found to lie
in the range $10^{-5}-10^{-6}$ in the Standard Model.

\acknowledgements

D. G. D.\ has been supported by a grant from the Commission
of the European Communities, under the TMR programme (Contract
N$^\circ$ ERBFMBICT961548). This work has been funded by
CICYT (Spain) under grant No. AEN-96-1718.

\begin{figure}[htbp]
\begin{center}
\vspace{1cm}
\epsfig{file=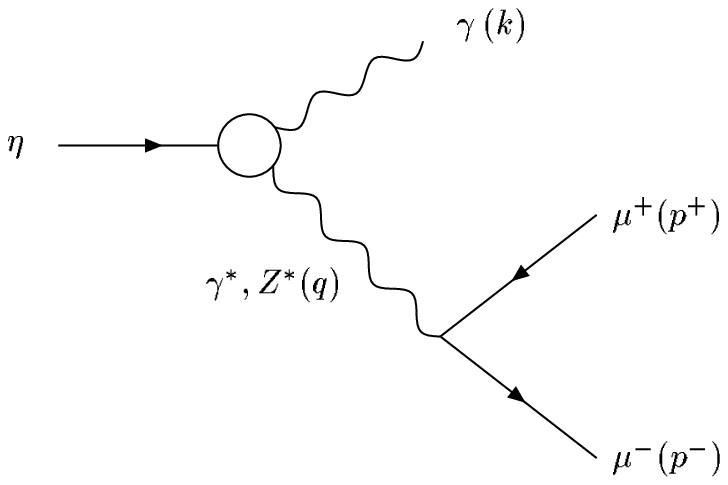}
\end{center}
\caption{Diagram for the $\eta\rightarrow\gamma\,\mu^+\mu^-$ decay. The
circle stands for the anomalous vertex.}
\end{figure}

\hfill

\begin{figure}[htbp] 
\begin{center} 
\epsfig{file=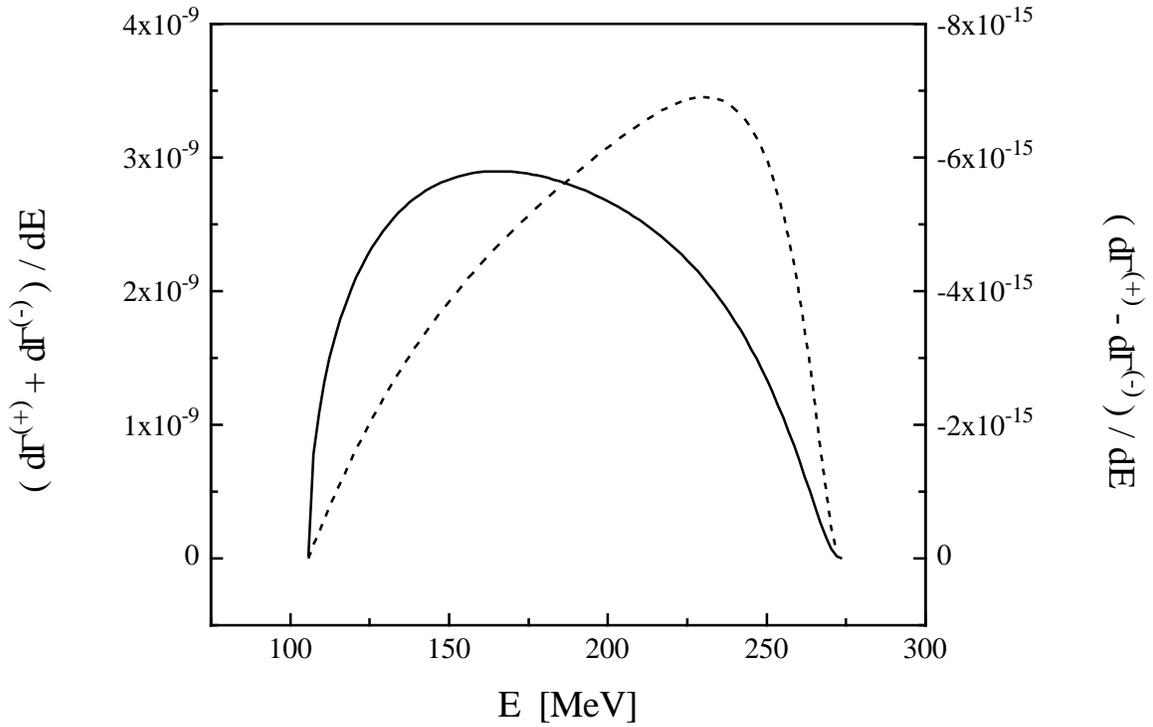}
\end{center} 
\caption{Differential decay rates for the process $\egmm$, in terms of the
energy of the final $\mu^+$. The solid line stands for the total width,
while the dashed one corresponds to the difference between rates
to opposite longitudinal $\mu^+$ polarizations. Notice the
different ordinate scales at both sides of the figure.}
\label{fig2} 
\end{figure} 
 
\end{document}